\documentclass[]{aa}
\usepackage{graphicx}
\usepackage{epsfig}
\usepackage{latexsym}
\usepackage{txfonts}
\usepackage{natbib}
\begin{document}
\renewcommand{\labelitemi}{-}
%title & authors
\title{The influence of albedo on the size of hard X-ray flare sources}
\author{M. Battaglia 
  \and E. P. Kontar \and I. G. Hannah }
\institute{School of Physics and Astronomy, University of Glasgow, Glasgow G12 8QQ, UK}
\offprints{M. Battaglia, \email{mbattaglia@astro.gla.ac.uk}}
\date{Received /Accepted}

%abstract
\abstract
%context
{Hard X-rays from solar flares are an important diagnostic of particle acceleration and transport in the solar atmosphere. However, \textit{any} observed X-ray flux from on-disc sources is composed of direct emission plus Compton backscattered photons (albedo). This affects both the observed spectra and images as well as the physical quantities derived from them such as the spatial and spectral distributions of accelerated electrons or characteristics of the solar atmosphere (e.g. density).  }
%aims
 {We propose a new indirect method to measure albedo and to infer the directivity of X-rays in imaging using RHESSI data. We describe this method and demonstrate its application to a compact disc event observed with RHESSI.}
%methods
{Visibility forward fitting is used to determine the size (second moment) of a disc event observed by RHESSI as a function of energy. Using a Monte Carlo simulation code of photon transport in the chromosphere, maps for different degrees of downward directivity and true source sizes are computed. The resulting sizes from the simulated maps are compared with the sizes from the observations to find limits on the true source size and the directivity. }
%results
{The observed full width half maximum (FWHM) of the source varies in size between 7.4 arcsec and 9.1 arcsec with the maximum between 30 and 40 keV. Such behaviour is expected in the presence of albedo and is found in the simulations. The uncertainties in the data are not small enough to make unambiguous statements about the true source size and the directivity simultaneously. However, a source size smaller than 6 arcsec is improbable for modest directivities and the true source size is likely to be around 7 arcsec for small directivities. }
%conclusions
{While it is difficult to image the albedo patch directly, the effect of backscattered photons on the observed source size can be estimated. This is demonstrated here on observations for the first time. The increase in source size caused by albedo has to be accounted for when computing physical quantities that include the size as a parameter such as flare energetics. At the same time, the study of the albedo signature provides vital information about the directivity of X-rays and related electrons. }

%\titlerunning{}
\authorrunning{Battaglia et al.}

\maketitle

% Introduction
%--------------------------------------------------------

\section{Introduction} \label{Introduction}
X-ray emission from solar flares is an important diagnostic of electron acceleration and transport in the solar atmosphere. In the common flare scenario, particles are accelerated in the corona. They precipitate along the field lines of a magnetic loop to the dense chromosphere where they loose their energy in Coulomb collisions with the surrounding plasma and produce hard X-ray (HXR) bremsstrahlung radiation. However, the detailed physics of electron acceleration and propagation is a subject of active research. 
Many studies of the different aspects of the physics involve the determination of the spatial characteristics (position, size) over which processes in flares happen. An accurate measurement of the size of X-ray flare sources is crucial for the computation of quantities such as the total energy involved in a flare \citep[eg.][]{Ca80, St84, Sa05,Ve05,Ha08,Fi09} or densities of coronal sources \citep[eg.][]{Ba09,Kr08}.

The Reuven Ramaty High Energy Solar Spectroscopic Imager \citep[RHESSI,][]{Li02} is capable of observing X-rays from solar flares with high spectral and spatial resolution. This allows us to determine the positions and sizes of HXR flare sources to very high accuracy. \citet{As02} measured the position of X-ray sources as a function of energy, finding a decrease of the radial position and size  with increasing energy. \citet{Ko08} used the new analysis method of visibility forward fitting \citep{Hur02,Sc07} on a limb event, finding a clear decrease of radial position with energy. It is now possible to not only find the position but also the sizes of the sources parallel and perpendicular to the magnetic field as demonstrated by \citet{Koet10} with visibility forward fitting. The parallel sizes provide valuable information on the physics of electron transport in the chromosphere. In a target with steeply increasing density in which the electrons predominantly lose their energy in Coulomb collisions, the parallel size is expected to decrease with energy. The size perpendicular to the field lines contains information about the convergence of the magnetic field. Since higher energetic electrons penetrate deeper into the chromosphere, a decrease of the perpendicular size with energy is also expected for a converging magnetic field. Such behaviour was found by \citet{Koet10} in a RHESSI limb flare. 

RHESSI's highest angular resolution in theory is 2.3 arcsec \citep{Hur02}. However, typically observed source sizes (in terms of FWHM) are often in the range of 10 arcsec for coronal sources \citep[eg.][]{Kr08} and microflare loops \citep{Ha08} and around 3-7 arcsec for footpoints \citep{De09}. In comparison, measurements of flare ribbons in EUV \citep[eg.][]{Wa07,Fl03} and white light \citep{Is07,Hu06} find much smaller structures on the scales of 1 arcsec. One common explanation for this is that RHESSI might not fully resolve those smaller structures. 

A more likely explanation is X-ray albedo: downward directed photons can be Compton backscattered in the photosphere and add to the observed X-ray flux. The influence of this on HXR spectra was first investigated by \citet{To72} and \citet{Sa73}. \citet{Bai78} and more recently \citet{Zh04} and \citet{Ko06} showed that this albedo flux can account for up to 40 \% of the detected flux in the energy range between 30 and 50 keV even in isotropic sources. The backscattered component originates from a large area compared to the size scale of typical flare sources. Thus the brightness of this albedo-patch is rather faint and direct imaging is difficult \citep{Sc02}. However, \citet{Ko10} recently found how the albedo patch will affect the measured size of the source even if direct imaging is not possible. Using Monte Carlo simulations of photon transport they calculated the expected source sizes and positions of X-ray sources on the solar disc, taking into account albedo. They found that the observed size depends on the location on the solar disc, the spectral hardness and the downward directivity (ratio of downward to upward photon flux $\alpha=I_{down}/I_{up}$). Furthermore, the size as a function of energy increases with increasing energy, peaking at around 40 keV, before decreasing. 
However, albedo is not just an unwanted effect that has to be corrected for but can be used to study the directivity of X-ray emission in solar flares. From the knowledge of the X-ray directivity, information on the directionality of the accelerated electron distribution can be gained as demonstrated by \citep{Ko06a}. Currently, this is the only available diagnostic of downward electron beaming, except for contradictive measurements of polarization \citep[see eg.][]{Su06,Bo06}. Numerical modelling \citep[eg.][]{Le83} suggests values of $\alpha$ in the range of 1-2. Recent observations of the X-ray directivity indicate values of 0.3-3 \citep{Ka07} and around 1 \citep{Ko06a}.

None of the previous studies on directivity and electron beaming used imaging observations to measure the directivity. Since the albedo patch is faint it is difficult to image directly. Furthermore it is hard to interpret different source contributions correctly in single images and to distinguish potential extended primary sources from albedo. In this work we discuss an indirect method, based on imaging spectroscopy of the moments of the HXR spatial distribution. Visibility forward fitting is used to determine the source size (second moment of the spatial X-ray flux distribution) as a function of energy. Since albedo is energy dependent, this should be reflected in the observed size. The method is applied to observations of an on-disc flare, as well as simulated photon maps. The Monte Carlo code developed by \citet{Ko10} was used to simulate photon maps for different true source sizes and degrees of directivity. From those maps, X-ray visibilities were created and fitted using visibility forward fitting the same way as the observed data was analysed. Since albedo is energy dependent with a peak around 40 keV the observed size of the source as a function of energy will follow a distinct pattern. Comparison of this observed pattern with predictions from the simulations makes it possible to find a measure for the true source size and the directivity.

In Section \ref{method} we describe the new method. Section \ref{modelling} describes how simulations of photon transport were performed for comparison with the observations. In Section \ref{observations} we apply the proposed method to observations of an actual RHESSI flare followed by a discussion (Section \ref{discussion}).
\section{Method}\label{method}
For an X-ray source of brightness $I(x,y)$ $[photons$ $cm^{-2}$ $arcsec^{-2}]$, the source size can be obtained from the second moment of the brightness distribution:
\begin{equation} \label{smoments}
x_{size}^2=\frac{\int_Sx^2I(x,y) \mathrm{d}x\mathrm{d}y}{\int_SI(x,y) \mathrm{d}x\mathrm{d}y}, \quad
y_{size}^2=\frac{\int_Sy^2I(x,y) \mathrm{d}x\mathrm{d}y}{\int_SI(x,y) \mathrm{d}x\mathrm{d}y},
\end{equation}
where we integrate over the whole emission area S. The source size (FWHM) is then given as $FWHM_{x,y}=2\sqrt{2\ln 2}(x_{size}/y_{size})$. In their simulations, \citet{Ko10} computed the moments directly from the simulated photon maps. For imaging observations of true flares with RHESSI, this is impractical for various reasons (background, CLEAN side lobes, artefacts of the imaging algorithm). We therefore use the method of visibility forward fitting \citep{Hur02,Sc07} to determine the FWHM of the source. Previous studies demonstrated the usefulness of visibilities to measure the moments of the HXR-distribution (zeroth moment: flux, \citet{Em03, Ba06}, first moment: position, \citet{Ko08}, second moment: size, \citet{Koet10,Xu08}). Visibility forward fitting is also ideally suited to look for albedo because it emphasises large scale structures in a source. This can be shown as follows. 
The second moment of the spatial X-ray distribution as defined in Eq. \ref{smoments} relates directly to the size found in visibility forward fitting. The X-ray visibilities of the source are the spatial Fourier transforms of the photon flux \citep{Hur02}:
\begin{equation} \label{xflux}
V(u,v)=\int_x\int_yI(x,y)e^{2\pi \mathrm{i}(ux+vy)}\mathrm{d}x\mathrm{d}y 
\end{equation} or in vector form 
\begin{equation} \label{xflux2}
V(\vec k)=\int_S I(\vec r)e^{2\pi \mathrm{i}\vec k \vec r}\mathrm{d}\vec r,
\end{equation}
where $\vec k=(u,v), \vec r=(x,y)$.
The total, spatially integrated X-ray flux is simply represented by $V(\vec k)$ at $\vec k =0$. Indeed $V(\vec k=0)=\int_SI(\vec r)\mathrm{d}\vec r$ as can be seen from Eqs. \ref{xflux} and \ref{xflux2}.
The size of the source is directly related to the second order derivative of Eq. \ref{xflux} at $\vec k=0$:
\begin{eqnarray} \label{secder}
\left.\frac{\partial^2 V(\vec k)}{\partial u^2}\right|_{u=0}&=&(2\pi \mathrm{i})^2\int_S x^2  I(\vec r) \mathrm{d\vec r} \\
\left.\frac{\partial^2 V(\vec k)}{\partial v^2}\right|_{v=0}&=&(2\pi \mathrm{i})^2\int_S y^2  I(\vec r) \mathrm{d\vec r} \label{secder1}.
\end{eqnarray}
Using Eqs. \ref{secder} and \ref{secder1} in Eq. \ref{smoments} the size of the HXR sources from X-ray visibilities is:
\begin{eqnarray} \label{sder}
x_{size}^2&=&\frac{\int_Sx^2I(x,y) \mathrm{d}x\mathrm{d}y}{\int_SI(x,y) \mathrm{d}x\mathrm{d}y}=\left. \frac{1}{(2\pi \mathrm{i})^2}\frac{1}{V(\vec k=0)}\frac{\partial^2 V}{\partial u^2}\right|_{\vec k=0} \\
y_{size}^2&=&\frac{\int_Sy^2I(x,y) \mathrm{d}x\mathrm{d}y}{\int_SI(x,y) \mathrm{d}x\mathrm{d}y}=\left.\frac{1}{(2\pi \mathrm{i})^2}\frac{1}{V(\vec k=0)}\frac{\partial^2 V}{\partial v^2}\right|_{\vec k=0}. \label{sder1}
\end{eqnarray}
Let us assume a Gaussian source $I(x,y)=\frac{I_0}{2\pi \sigma^2}e^{(-(x^2+y^2)/2\sigma^2)}$, where the centroid is at $x=0,y=0$. The size of the source is represented by the Gaussian sigma. 
The Fourier transform of a Gaussian is also a Gaussian: $V(u,v)\sim e^{-2\pi^2\sigma^2(u^2+v^2)}$.
Substituting this in Eqs. \ref{sder} and \ref{sder1} one gets the sizes of the HXR sources: $x_{size}^2=\sigma^2$ and  $y_{size}^2=\sigma^2$. Thus, fitting Gaussian sources to RHESSI visibilities naturally emphasises the large-scale structures of any flare source and the resulting size is directly related to the second moment of the HXR emission.
\subsection{Comparison of observations and modelling} \label{modelling}
\begin{figure}
%\resizebox{\hsize}{!}{\includegraphics{sourceim.eps}}
\includegraphics[height=90mm]{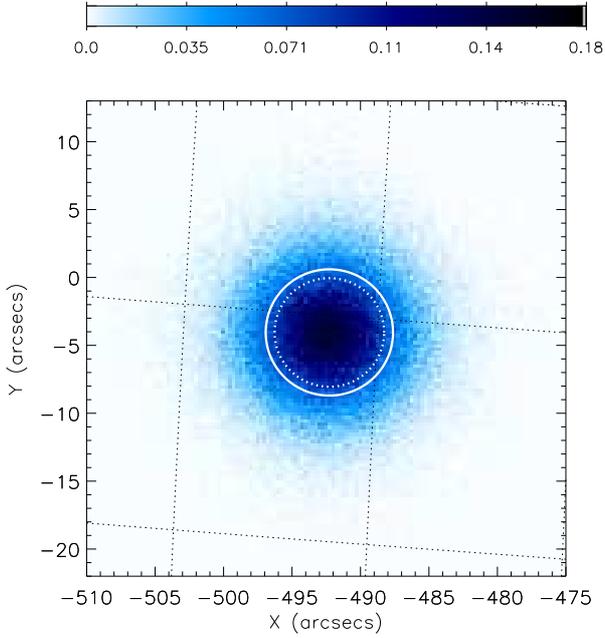}
\caption {Simulated photon map I(x,y) for true source of size 8 arcsec FWHM \textit{(white dashed circle)} at energies 30-45 keV. The \textit{white line} indicates the FWHM of a circular Gaussian source that was forward fitted to visibilities corresponding to the simulated map.}
\label{mapfig}
\end{figure}

To find a measure of the true source size and constrain the directivity, measurements of the observed FWHM have to be compared with predictions from the code developed by \citet{Ko10}.
The following steps are taken: (i) for a given true source size and downward directivity, a photon map of the total (direct and backscattered flux) is produced for a source at a heliocentric angle corresponding to the location of the observed flare (see Fig. \ref{mapfig}). 
(ii) The photon map is Fourier transformed to get the corresponding visibilities. (iii) Those visibilities are fitted using visibility forward fitting to find the FWHM, in the same way as for the real data. This size is directly comparable to the FWHM found for the data.
The simulated size depends on two free parameters, the true source size and the directivity. Thus one cannot infer either of those parameters from direct fitting. One can find one assuming the other and compare the resulting size behaviour to the observations, finding upper and lower limits of the directivity and the true source size.
\section{Observations} \label{observations}
In this section we present observations of a compact flare to which we applied the above described method.
\begin{figure}
%\resizebox{\hsize}{!}{\includegraphics{lcurvespec2.eps}}
\includegraphics[height=75mm]{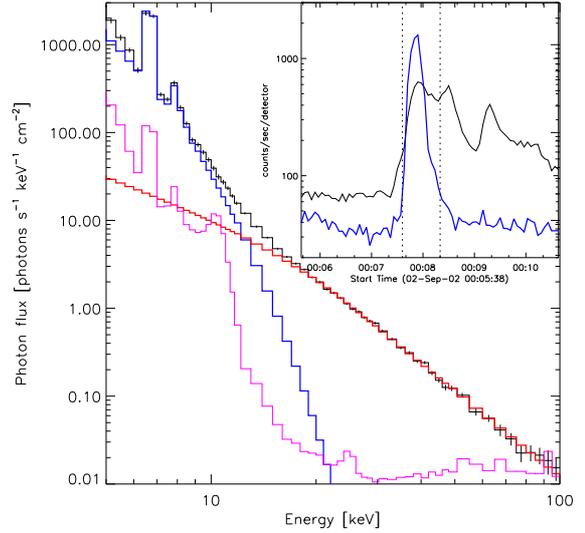}
\caption {Spectrum of the observed event, fitted with a thermal component (\textit{blue}) plus a thick target power-law model (\textit{red}). \textit{Purple} gives the background level. Inset right-hand corner: Lightcurve of the event at 6-12 keV (\textit{thin line}) and 25-50 keV (\textit{thick line}). The \textit{dashed} lines indicate the observed time interval. }
\label{spec}
\end{figure}
The flare presented here was a compact GOES C2.2 class event on the solar disc. It was observed on the 2nd September 2002 with peak time 00:08 UT at a position of E31N6. This corresponds to a heliocentric angle of $\theta=30\degr$ (or expressed in terms of the cosine: $\mu=cos(\theta)=0.86)$.
% cosine of the heliocentric angle of $\mu=cos(\theta)=0.86$.
The event is well suited for analysis by the moment based approach since it has a simple structure. Furthermore, the albedo effect is more pronounced close to the centre of the solar disc. Thus, a significant albedo contribution is expected. % $x=-492'', y=-4.6''$ 
Figure \ref{spec} (inset) shows the RHESSI lightcurve of the event at 6-12 keV and 25-50 keV. Full Sun spectroscopy was used to determine the shape of the spectrum (needed for modelling), as well as the energy range of the thermal and non-thermal emission. Figure \ref{spec} displays the spectrum fitted with a thermal plus thick target model with albedo correction for an isotropic source. We find that the spectrum is non-thermal above about 16 keV with an electron spectral index $\delta=4.1$ {corresponding to a photon spectral index $\gamma=3.1$ in a thick target.
\subsection{Imaging}   
\begin{figure*} %[!]
\begin{minipage}[l]{0.9\textwidth}
\includegraphics[width=160mm, height=50mm]{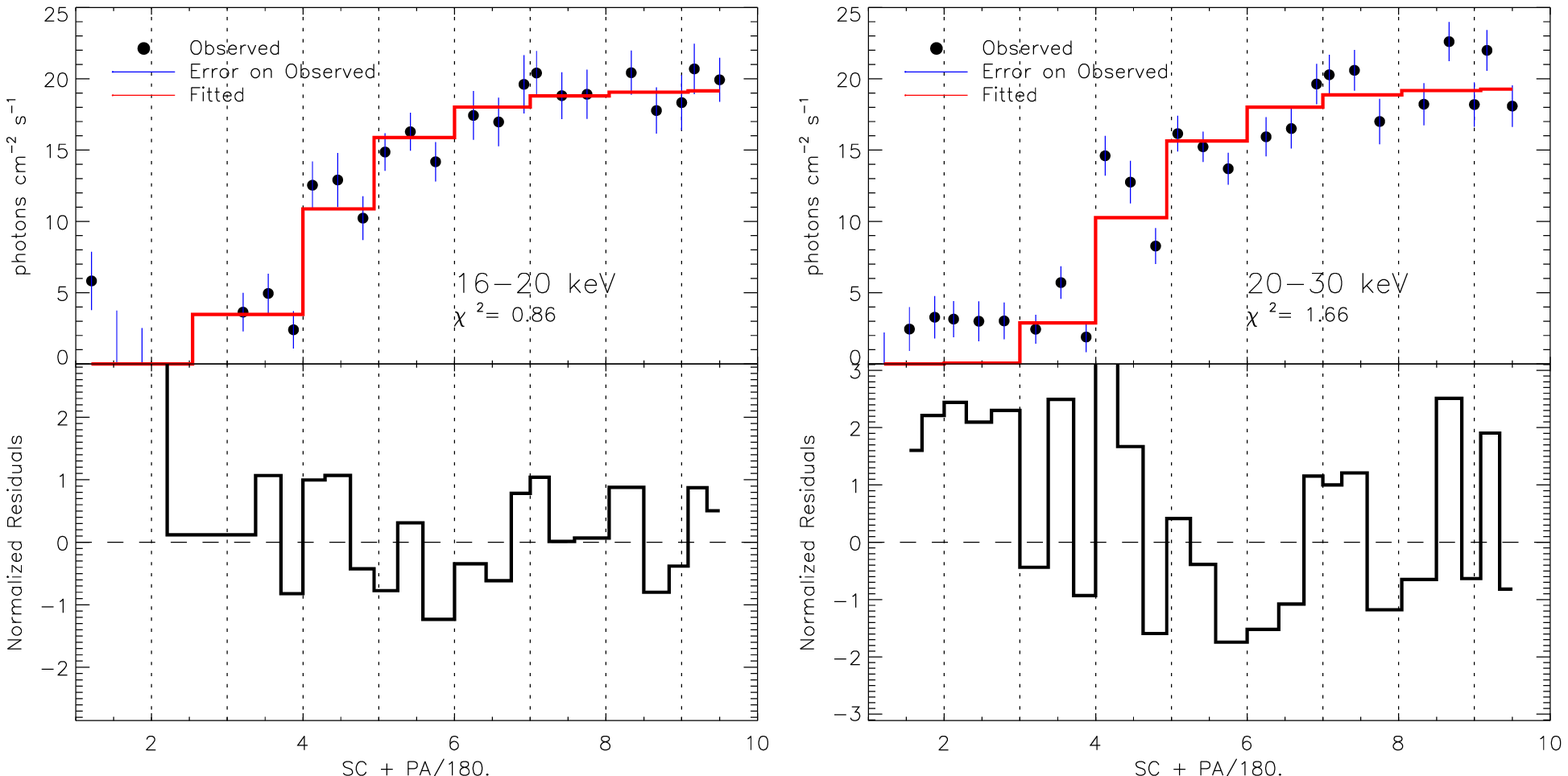}
\end{minipage}
\begin{minipage}[l]{0.9\textwidth}
\includegraphics[width=160mm, height=50mm]{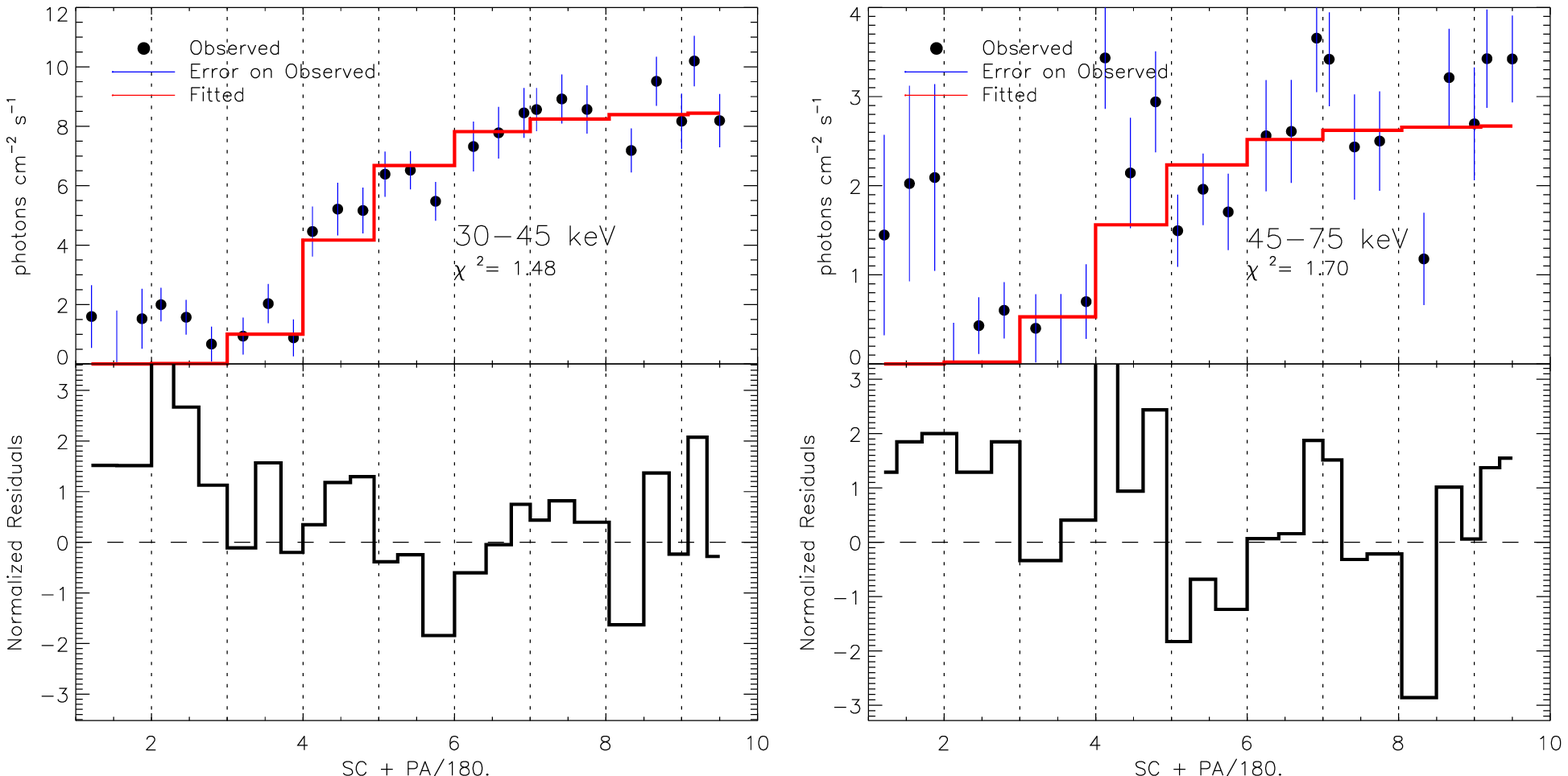}
\end{minipage}
\caption{Visibility fits for the four energy bands. The top half in each panel shows the measured visibility amplitudes \textit{(black dots)} with statistical errors (\textit{blue lines}) and the model fitted to the visibilities (\textit{red}). The bottom half in each panel gives the normalised residuals. }
\label{visfits}
\end{figure*}
\subsubsection{Visibility forward fitting}
Four energy bands (16-20, 20-30, 30-45, 45-75 keV) of the non-thermal emission (Fig. \ref{spec}) were used for the analysis. The energy bands were chosen in order to cover the energies with non-thermal emission, yet allowing for a high enough signal to noise ratio in each energy band. A time interval of 44 seconds was used, which spans the full HXR peak emission (dashed lines in the inset in Fig. \ref{spec}). The source was fitted with a single circular Gaussian model. The visibility fits are shown in Fig. \ref{visfits} for each of the four energy bands. The top half of each panel shows the measured visibility amplitudes with errors and the corresponding fit while the bottom half gives the normalised residuals. The visibilities suggest negligible modulation in the finest grid (1), and only small modulation in the second finest grid (2), suggesting a lack of detectable structure at the spatial scales of detectors 1 and 2. This gives a first lower limit of the source size between 3.9 and 6.8 arcsec.
Fits using all grids compared to using only grids 3-9 gave equal results within uncertainties, further confirming that the contribution of the finer grids is small. Also, the chosen number of roll bins (coverage in the uv-plane) does not affect the fitted size. While a large number of roll bins might be important for the measurement of higher moments (e.g. shape of the source), the uncertainties of the smaller moments (position,size), are larger than any effect a different set of roll bins has on the fitted parameters.
\subsubsection{Other imaging algorithms} \label{subpixon}
In addition to visibility forward fitting, CLEAN \citep{Hur02} and Pixon \citep{Me96} images were also generated for a better understanding of the source structure. Figure \ref{allmapfig} displays contour plots of the flare at different energy bands using these 3 imaging algorithms.
Pixon indicates a substructure in the 30-45 keV energy band. The extent of the separation of the two sources strongly depends on the Pixon-sensitivity (how sensitive Pixon reacts to small scale structures) and grids used (for the images in Fig. \ref{allmapfig}, grids 2-8 were used). It could be argued that these are two footpoints of a loop but it is not possible to make any unambiguous statements. The existence of two sources is confirmed to some extent by CLEAN. However, this is quite sensitive to the parameters used in the image reproduction. The substructure is apparent in CLEAN when uniform weighting (more emphasis is given to the finer grids) is used instead of natural weighting. On the other hand, it appears in images when using grids 2-8 with natural weighting and by changing the width of the CLEAN beam (a clean beam width factor of 2 was used for the generation of the images in Fig.~\ref{allmapfig}). In the highest energy band (45-75 keV), CLEAN is affected by noise and the 30 \% contour in Fig. \ref{allmapfig} outlines some of those artefacts. See Section \ref{resanddiss} for a more thorough discussion of how this might affect our results. 
\begin{figure*}
\resizebox{\hsize}{!}{\includegraphics{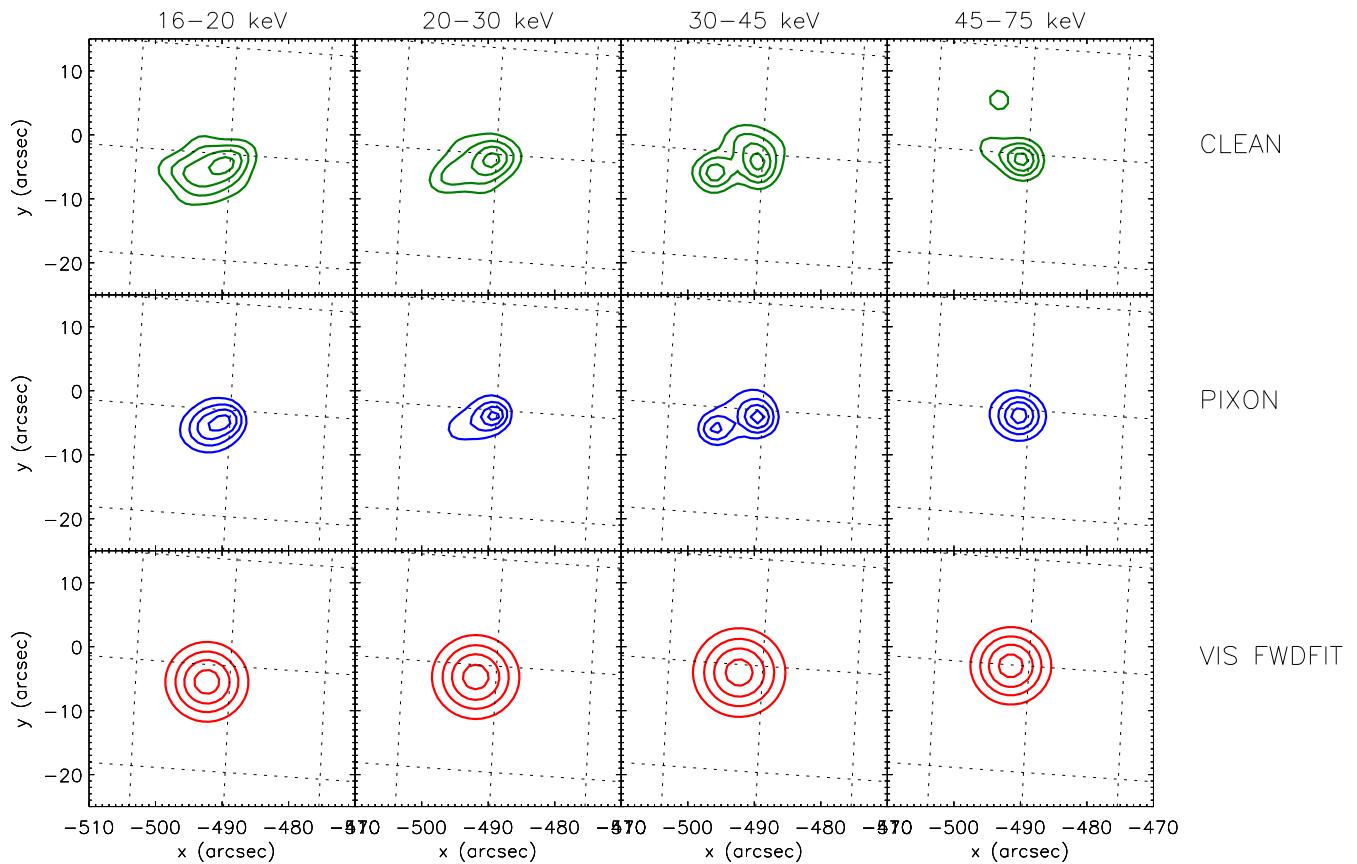}}
\caption {Contour maps (30, 50, 70, and 90 \% of maximum emission) from three different algorithms (top to bottom: CLEAN, Pixon, visibility forward fitting) and energy bands (left to right: 16-20 keV, 20-30 keV, 30-45 keV, 45-75 keV).}
\label{allmapfig}
\end{figure*}
\begin{figure}
\centering
\resizebox{\hsize}{!}{\includegraphics{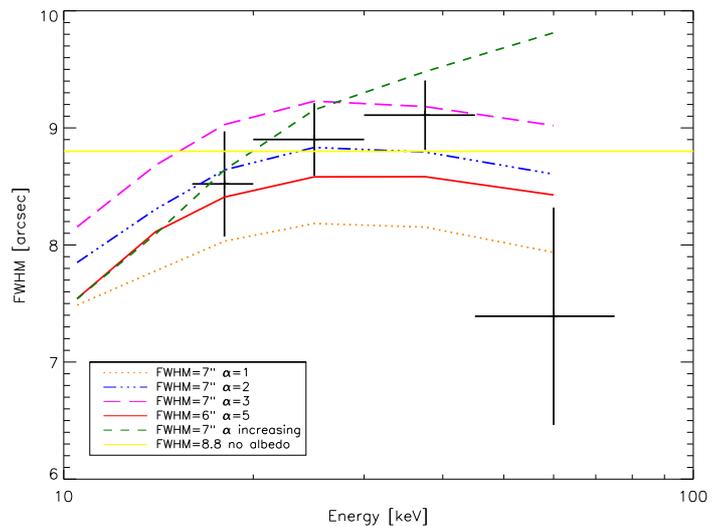}}
\caption {Observed FWHM as a function of energy (data points with error bars). FWHM from visibility forward fitting to simulated data for 3 primary source sizes and different degrees of directivity (see legend in the figure).}
\label{fwhmfig}
\end{figure}
\subsection{Results}
Figure \ref{fwhmfig} displays the observed source size found from visibility forward fitting (data points).
The FWHM was found to increase from 8.5$\pm$0.4 arcsec in the 16-20 keV energy band to 9.1 $\pm$  0.3 arcsec in the 30-45 keV energy band, then decreases to 7.4 $\pm$ 0.9 arcsec. 
This was compared to simulations as described in Section \ref{modelling}. For the simulations, a heliocentric angle $\mu$ between 0.8 and 0.9 was used and a source height of 1 Mm above the photosphere. Due to the large uncertainties in the data, no unambiguous statements about either directivity or true source size can be made. However, one can give a lower limit of the true source size, as well as an estimate of the directivity. Two source sizes were simulated (6 arcsec and 7 arcsec) for various degrees of directivity. The orange, blue and purple curves in Fig.~\ref{fwhmfig} indicate the expected size for a true source size of 7 arcsec and directivities $\alpha=1$, $\alpha=2$ and $\alpha=3$ respectively. The red curve gives the expected size for a true source size of 6 arcsec and directivity $\alpha=5$. This excludes a very small source since very large directivities would be needed to explain the observed size. The dashed green curve indicates a model for which the directivity is increasing with increasing energy as $\alpha=\epsilon/10$ where $\epsilon$ is the photon energy in keV. Such a behaviour would be expected in a collisional target since electrons with smaller energies are expected to have a higher number of collisions which will result in a larger spread of the pitch angle \citep[see eg.][]{Br72}. Finally, a hypothetical constant source size is indicated by the yellow line. Table \ref{prob} provides the probabilities that a chosen model is consistent with the data according to the $\chi^2$ test \citep{Pr92}.
 \begin{table}
\caption{Probabilities that a given model is consistent with the data according to the $\chi^2$ test.}
\begin{center}
\begin{tabular}{ll}
Model & Probability that model is\\ 
& consistent with data\\
\hline
FWHM=7'' $\alpha=1$ & 0.2 \%\\
FWHM=7'' $\alpha=2$ & 56 \%\\
FWHM=7'' $\alpha=3$ & 24 \%\\
FWHM=6'' $\alpha=5$ & 24 \%\\
FWHM=7'' $\alpha=\epsilon/10$ & 6 \%\\
FWHM=8.8'' no albedo & 42 \% \\
\hline
\end{tabular}
\end{center}
\label{prob}
\end{table}
On these grounds, a true source size of 7 arcsec in the case of isotropic albedo can be rejected confidently, as can the case of increasing directivity with energy. Previous observational studies found directivities around $\alpha=1$ to $2$ \citep{Ka07,Ko06a}. Provided that this is a reasonable assumption for the flare presented here, the true source size is then most likely of the order of 7 arcsec for a directivity of $\alpha=2$.
\section{Discussion} \label{discussion}
The method described here used the energy dependence of albedo and the ability of visibility forward fitting to get an indirect measure of albedo in imaging. This is an advance compared to previous attempts in which albedo was only tried to be imaged in one energy bands. The limitation of the presented method is clearly the large uncertainties in the measured size that do not allow an unambiguous fit of any model to the observations. However, they do provide important constraints on the source size and directivity. Further, the morphology of the analysed source has to be investigated carefully. Those points are addressed in the following paragraphs.
\subsection{Source morphology} \label{resanddiss}
While we interpret and fit the observed flare as a single source, it cannot fully be excluded that the source is in reality a double-source in the 30-45 keV energy band as hinted by Pixon (see Sect. \ref{subpixon} and Fig. \ref{allmapfig}), representing the footpoints of a flare loop. Since CLEAN finds indications of this structure, as well, it is unlikely that it is purely over-resolution caused by Pixon. However, visibility forward fitting was unsuccessful in fitting two sources reliably. Therefore, the single Gaussian model represents a good fit to the data but the spatial morphology in the third energy band is expected to have an influence on the fitted source size. As can be seen in Fig. \ref{fwhmfig}, the size in the third energy band is somewhat larger than what would be expected from the general trend of increasing and decreasing size. While this might enhance the effect of increasing and decreasing source size, the underlying cause is likely to be albedo. At the same time, the size in the fourth energy band (45-75 keV) is smaller than expected from any of the models. This could be caused by the smaller signal to noise ratio (weaker source) in this energy band that affects the result of visibility forward fitting (see Fig. \ref{visfits}). This is also reflected in the uncertainty of the size in this energy band.

In principle, the observed source could also be a very dense coronal source, as was found in other RHESSI observations \citep[e.g.][]{Ve05a}. This scenario can, however, be ruled out as follows. The spectrum indicates a rather weak thermal component with emission measure of $7.4\times 10^{46}\mathrm{cm^{-3}}$. Assuming a true source FWHM of 7 arcsec ($5.1\times 10^8\mathrm{cm}$), the density of the source would be $2.8\times 10^{10}\mathrm{cm^{-3}}$. This corresponds to a stopping depth of $3.2\times 10^9 \mathrm{cm}$ for an electron with energy 30 keV which is much larger than the extent of the source. The electrons will therefore be stopped much lower down in the atmosphere, indicating that the HXR emission is indeed chromospheric and not coronal. 
\subsection{Directivity and primary source size}
The FWHM of the observed source is found to increase and decrease with energy with a peak between 30 keV and 40 keV. This is contrary to what one would expect in the classical thick target model \citep{Br71}. Collisional energy losses would lead to a smaller extent of the source along the direction of propagation (parallel to the magnetic field) as a function of energy. Since electrons with higher energies penetrate deeper into the target, the extent of the source perpendicular to the magnetic field is expected to decrease as a function of energy in a converging magnetic loop. An overall decrease of the source size with energy would therefore be expected. Such behaviour was observed by \citet{Ko08} and \citet{Koet10} in a limb event for which albedo has little effect. 
 
The most obvious explanation for the observed behaviour presented in this paper is albedo. The contribution of backscattered flux is negligible below 10 keV due to photoelectric absorption. Photons with energies larger than 100 keV penetrate deeply into the atmosphere and are lost to the observer. The albedo contribution is therefore largest between 10 keV and 100 keV (the energy range observed here) with a maximum around 30 - 40 keV - where the maximum in source size is measured. The observed increase and decrease of the size is therefore consistent with the expected increase and decrease of the backscattered component. The observed size depends on two main parameters, the true source size, as well as the directivity. A third quantity that has to be considered is the height of the source above the photosphere. In the simulations we assumed a height of 1 Mm. This is a reasonable assumption in a standard atmosphere. Values of this order were also found observationally \citep{Sa10,Koet10, As02}. Since we cannot directly fit the true source size nor the directivity, we infer one assuming the other and compare the resulting size behaviour to the observations. As shown in Fig. \ref{fwhmfig}, the data are consistent with a true source size of 7 arcsec for a downward directivity of $\alpha=2$, a value that is consistent with previous observations \citep[][]{Ka74, Ves87, Ka07,Ko06a}. On the other hand, the inferred true source size is rather large for a footpoint, compared with previous measurements \citep[e.g.][]{De09}. Even for a large directivity such as $\alpha=5$ the true source size could be of the order of 6 arcsec. It is conceivable that the true source is smaller if the velocity distribution of the X-ray producing electrons is not beam-shaped but rather takes the form of a ``pancake'' distribution. In this case, the downward directivity would be the same, but with a higher contribution of reflected photon flux from the sides, leading to a larger albedo patch and therefore larger source size. 
\subsection{Direct and indirect imaging}
We chose an indirect approach to measure the albedo because the albedo patch is faint compared to the main source and is difficult to image directly. In CLEAN, any albedo component will most likely be masked by side-lobes of the main source. It has been suggested that it might be possible to image the albedo source directly using Pixon. We investigated this possibility using simulated flare data. Starting from simulated photon maps of a Gaussian source with albedo (no background), calibrated event lists were computed. The RHESSI software was then used to reconstruct Pixon images from the calibrated event lists. For a small primary source (2 arcsec) and very high countrates, Pixon seems to image the albedo component in the form of ``wings'' at larger distances from the source centre. However, for weaker and larger sources (comparable with the data presented here) this does not seem to work. A primary source with 8 arcsec FWHM was simulated both with and without albedo component. Figure \ref{pixon} illustrates the radial profile of the simulated source (summed over the y-axis) compared to the radial profile of the reconstructed Pixon image for the case of isotropic albedo ($\alpha=1$, left) and without albedo (i.e. pure Gaussian $\alpha=0$, right). The Pixon image depends strongly on certain parameters used for the image reconstruction, in particular the Pixon sensitivity. Two main points should be noted. First, in the albedo case Pixon finds ``wings'' at larger distances from the source centre that can be interpreted as the albedo component. However, in the $\alpha=0$ scenario (no albedo), Pixon indicates some emission at larger distances from the source centre that was not existent in the initial simulated image and must be artefacts from the image reproduction. In real observations it will therefore be difficult to judge whether such ``wings'' are truly albedo or just imaging artefacts. Secondly, as has been known previously, Pixon has a tendency to over-resolve sources. This is clearly demonstrated here where a non-existent substructure is found in a source that is a single Gaussian. 
\begin{figure*}
%\resizebox{\hsize}{!}{\includegraphics{sourceim.eps}}
\includegraphics[width=160mm, height=80mm]{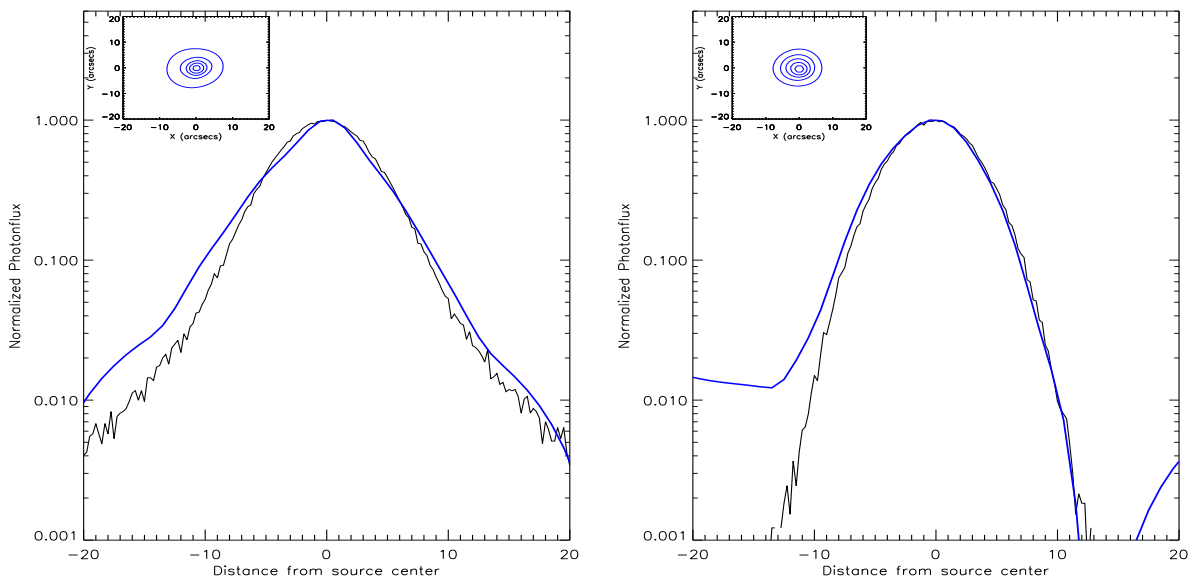}
\caption {Radial profiles (summed over the y-direction) of simulated photon maps including albedo (directivity $\alpha=1$, \textit{left}) and without albedo (\textit{right}). The \textit{thick blue line} indicates the radial profile through the Pixon images computed from the simulations. The inset shows contour plots of the Pixon images.}
\label{pixon}
\end{figure*}
Another possible approach is to fit the albedo patch directly using visibility forward fitting, i.e. to fit two sources one of which is the (small) main source, the other the (large) albedo patch. In one of the simulations it was possible to fit one smaller source (the size of the initial Gaussian) plus a larger Gaussian source ($\sim$ 15 arcsec FWHM) as the albedo component. Again, this is dependent on good count statistics and was not successfully applied to the observed flare data. 

\section{Conclusions} \label{conclusions}
In this work we discuss a new method to measure albedo in imaging. Since direct imaging of the faint albedo source is difficult, an indirect approach via imaging spectroscopy of the size (second moment of the X-ray source distribution) is better suited. The method combines the energy dependence of the albedo with the ability of visibility forward fitting to measure large scale sources structures. This is a step forward compared to previous studies which only used one energy band to image albedo \citep[eg.][]{Sc02}. We emphasise that different energy bands have to be analysed simultaneously to distinguish potential extended primary sources from albedo. 

The observed size versus energy dependence contradicts the expected behaviour in a classical thick target model. It can most easily be attributed to albedo. Given the albedo interpretation of the source, we find that the true source size is likely to be of the order of 7 arcsec for realistic values of the directivity in the order of $\alpha =2$ which would be consistent with earlier observational studies of anistropy of albedo \citep[]{Ka74, Ves87, Ka07,Ko06}. A model involving increasing directivity with energy can be ruled out by the data. 

Although the proposed method is limited to simple source morphologies at the moment, it clearly demonstrates the influence albedo has on observed X-ray sources and the derived sizes, a fact that has to be considered when inferring physical properties of flares. Further, it opens the way to determining the directivity of X-rays and inferring the directionality of accelerated electrons not only from spectra, but from imaging, for the first time. If the uncertainties could be reduced by a factor of 3, a confident rejection of a wider class of beam transport models would be possible. This might be achieved in a (future) flare with higher count rates or/and by further improving RHESSI visibility analysis, as well as in measurements made by future instruments.

\mdseries
\begin{acknowledgements}
We thank Brian Dennis for insightful comments and suggestions, Gordon Hurford for helpful discussions and Natasha Jeffrey for comments on the text. This work is supported by the Leverhulme Trust (M.B., E.P.K.) an STFC rolling grant (I.G.H., E.P.K.)
and STFC Advanced Fellowship (E.P.K.). Financial support
by the European Commission through the SOLAIRE Network
(MTRN-CT-2006-035484) is gratefully acknowledged. The work has benefited from an international team grant from ISSI Bern, Switzerland.
\end{acknowledgements}

\bibliographystyle{aa}
\bibliography{mybib}

\end{document}